\documentclass{article}
\usepackage{ijcai21}

\usepackage{times}
\usepackage{soul}
\usepackage{url}
\usepackage[hidelinks]{hyperref}
\usepackage[utf8]{inputenc}
\usepackage[small]{caption}
\usepackage{graphicx}
\usepackage{amsmath}
\usepackage{amsthm}
\usepackage{booktabs}
\usepackage{algorithm}
\usepackage{algorithmic}

\makeatletter
\newcommand*\titleheader[1]{\gdef\@titleheader{#1}}
\AtBeginDocument{%
\let\st@red@title\@title
\def\@title{%
\bgroup\normalfont\large\centering\@titleheader\par\egroup
\vskip1.5em\st@red@title}
}
\makeatother

\date{}

\title{\Large \bf CybORG: A Gym for the Development of Autonomous Cyber Agents}

\titleheader{\emph{IJCAI-21 1\textsuperscript{st} International Workshop on Adaptive Cyber Defense}}

\author{
Maxwell Standen\and  
Martin Lucas \and 
David Bowman \and 
Toby J. Richer \and 
Junae Kim \And 
Damian Marriott\\
\affiliations
Defence Science and Technology Group\\
\emails
\{max.standen, martin.lucas,  david.bowman, toby.richer, junae.kim, damian.marriott\}@dst.defence.gov.au}

\begin{document}
\maketitle
\hyphenation{CybORG}
\begin{abstract}
Autonomous Cyber Operations (ACO) involves the development of blue team (defender) and red team (attacker) decision-making agents in adversarial scenarios. To support the application of machine learning algorithms to solve this problem, and to encourage researchers in this field to attend to problems in the ACO setting, we introduce CybORG, a work-in-progress gym for ACO research. CybORG features a simulation and emulation environment with a common interface to facilitate the rapid training of autonomous agents that can then be tested on real-world systems. Initial testing demonstrates the feasibility of this approach.
\end{abstract}
\section{Background}
Autonomous Cyber Operations (ACO) is concerned with the defence of computer systems and networks through autonomous decision-making and action. It is particularly needed where deploying security experts to cover every network and location is becoming increasingly untenable, and where systems cannot be reliably accessed by human defenders, either due to unreliable communication channels or adversary action.

The ACO domain is challenging to develop artificial intelligence (AI) approaches for as it combines hard problems from other domains of AI research. Like game AI, it is adversarial: the effectiveness of a defensive cyber agent is determined by its ability to respond to an adversary. Like autonomous robotics, ACO is affected by the `reality gap' \cite{Ibarz_2021}, as simulations of an environment will abstract away information that could be critical to an agent's effectiveness. A further issue for the ACO domain is that the environment and action set change as cyber security research progresses, which is far more rapidly than either of the domains discussed above. 

The requirement to handle the varying actions of an adversary, in a complex environment, precludes the use of static data sets to learn ACO behaviour. A tool for learning in adversarial environments is an AI Gym. AI Gyms such as the one developed by OpenAI implement reinforcement learning (RL) through direct interaction with a simulation of the problem. 
 A path to addressing the `reality gap', used in \cite{7759424}, is to combine learning on simulations with testing in a real environment. In this case, the bulk of learning is conducted on simulated systems. Successful agents are transferred to the real system to firstly validate their effectiveness, and secondly to refine the simulation. 

We believe that AI Gyms, that can be validated and refined through experiments in real-world environments, are the most promising tool for training effective intelligent cyber agents. This paper presents the Cyber Operations Research Gym (CybORG), an environment designed to facilitate the development of ACO agents through this two-stage process. 
\section{Related Work}
\begin{table*}[t] \centering
\resizebox{\textwidth}{!}{ 
\begin{tabular}{|l|l|l|l|l|l|l|l|}
\hline
& Simulation & Emulation & Scalable & Flexible & Efficient & Adversarial CO & Designed for RL \\ \hline
DETERlab~\cite{5655108} & No & Yes & Low & Yes & Low & No & No \\ \hline
VINE~\cite{Eskridge:2015:VCE:2808475.2808486} & No & Yes & Med & Yes & Med & No & No \\ \hline
SmallWorld~\cite{FURFARO2018791} & No & Yes & Med & Yes & Med & No & No \\ \hline
BRAWL~\cite{BRAWL} & No & Yes & Med & Windows & Med & No & Limited \\ \hline
Galaxy~\cite{220241} & No & Yes & Low & Debian-based & High & Limited & Yes \\ \hline
Insight~\cite{Futoransky2009SimulatingCF} & Yes & No & High & Yes & Med & No & No \\ \hline
CANDLES~\cite{Rush:2015:CAN:2739482.2768429} & Yes & No & High & Yes & High & Limited & Yes \\ \hline
Pentesting Simulations~\cite{niculae_2018,JThesis} & Yes & No & High & Yes & Med & Limited & Yes \\ \hline
CyAMS~\cite{7795375} & Yes & Yes & High & Yes & High & No & No \\ \hline
CyberBattleSim~\cite{msft:cyberbattlesim} & Yes & No & High & Yes & High & Limited & Yes \\ \hline
FARLAND~\cite{farland} & Yes & Yes & High & Networks only & High & Yes & Yes \\ \hline
\bf CybORG & \bf Yes & \bf Yes & \bf High & \bf Yes & \bf High & \bf Yes & \bf Yes \\ \hline
\end{tabular}
}
\caption{Existing Environments and CybORG Design vs ACO Research Requirements}
\label{Req Table}
\end{table*}
There are a growing number of cyber security environments designed for experimentation. A summary of several environments, with an assessment of how they fit our requirements, can be found in Table \ref{Req Table}. 

DETERlab~\cite{5655108} is a specialised cyber security experimentation environment based on EMUlab~\cite{Hibler:2008:LVE:1404014.1404023}. It supports cyber security experimentation through the emulation of hosts and networks. As it relies on local hardware, DETERlab has limited maximum network size and takes a significant amount of time to reset or reconfigure. VINE~\cite{Eskridge:2015:VCE:2808475.2808486}, SmallWorld~\cite{FURFARO2018791} and BRAWL~\cite{BRAWL} leverage cloud-based Infrastructure as a Service (IaaS) and virtualisation frameworks such as OpenStack to emulate larger enterprise-like networks. These tools can simulate users acting on a host which are designed to generate human-like activity for the purpose of experimentation with different tools. BRAWL can make use of CALDERA~\cite{Applebaum:2016:IAR:2991079.2991111,CalderaPOMDP} as a red team agent to produce a more realistic experimentation testbed.

GALAXY~\cite{220241} is an emulated cyber security environment. It has been used for research into evolutionary algorithms, training red agents to blend in with normal user traffic. It is currently restricted to a single VM per physical machine, but can use VM snapshotting to perform fast resets of the environment. 

Insight~\cite{Futoransky2009SimulatingCF} is a highly scalable system for simulating networks for the training of offensive agents. It can simulate hundreds of hosts on a single computer, but does not support the training of defensive agents.

CANDLES~\cite{Rush:2015:CAN:2739482.2768429} leverages a high speed network security simulation to coevolve adversarial blue and red agents using an evolutionary algorithm. CyAMS~\cite{7795375} uses a combined emulation and simulation environment. CyAMS simulates a cyber security environment using a Finite State Machine (FSM). CyAMS also features an emulation environment and compares the simulation with the emulation using a malware propagation scenario to demonstrate the fidelity of the simulation.

CyberBattleSim~\cite{msft:cyberbattlesim} focuses on a red agent performing post breach lateral movement in a windows enterprise environment. The simulation is node-based which allows for a highly flexible state and action space at the cost of a high level observation space which lacks information on second order effects. The simulation allows the use of a fixed defender agent but does not support the training of the defender agent. CyberBattleSim is a pure simulation and is not directly connected to an emulation.

FARLAND~\cite{farland} is another framework that is designed for the training of agents in simulation and testing of agents in emulation. It provides some useful extensions to CybORG's current functionality, such as probabilistic representations of state and support for adversarial and deceptive red agents. However, FARLAND is designed for network traffic exploitation and defence, rather than host-based exploitation and defence. It could serve as a useful complement to CybORG for scenarios with more of a focus on the properties of the network.

Each of these training environments fulfill some of our requirements. We want to train red and blue agents that can interact realistically with a variety of hosts. We require the ability to rapidly train these agents in simulated environments and to then test and validate them in emulated environments. To our knowledge, all of these features are not available in any single environment described above. In the next section, we present the design of our system which targets these requirements.
\section{Overview of CybORG}

CybORG is designed to implement a range of scenarios at differing levels of fidelity through a highly modular design. Once a scenario is fully defined within CybORG, an agent can interact with that scenario modelled in a finite state machine (defined here as simulation) or fully implemented in virtual infrastructure (defined here as emulation). 

The CybORG tool generates a scenario based on a pre-generated description, initialises a set of agents to perform roles within that scenario, then implements their actions and assesses their effectiveness. CybORG then runs the scenario in a series of discrete steps. At each step, each agent selects an action to perform from its action space. An agent’s action space is a subset of the overall available set of actions, dependent on the role of the agent and defined as part of the scenario. Once each agent has selected an action, it is performed. The agent receives an observation of the updated state of the scenario. The scenario is run until it reaches its termination condition -- either a limit on the number of steps, or the achievement of an agent's goal. At this point, the agents receive any final information on the state of the scenario. The scenario can be reset for further training or testing.

To implement CybORG's two levels of fidelity, the scenarios and actions used by CybORG are defined at two levels. The scenario definition contains the required information to simulate the scenario. It also contains the system images required to emulate the scenario. Each action used by CybORG is defined twice. For simulation, each action is defined as a state transition. For emulation, each action defines a command (with appropriate parameters) that can be executed to achieve the desired effect. 

\section{Design}

\subsection{Scenarios}

A CybORG scenario defines the `game' that agents are aiming to solve or compete in. This scenario defines what agents exist, what actions they may perform, what information they begin the game with, and how their reward is calculated. It defines the configuration of each host and the network connections between them.

Deploying a particular scenario involves specifying whether it will be simulated or emulated and providing a scenario description file in the data-serialization format YAML. The scenario file includes details for configuring the environment including hosts, networking and subnet information, and the set of actions available to red and blue agents. The scenario files are deliberately simple, requiring a minimum of information and employing many default behaviours to reduce the burden on users when creating files.

The supported host configurations and actions are specified in separate `Images' and `Actions' YAML files respectively. The host file contains information on operating system, services, processes, users and other system information. It also contains an identifier for a deployable image to be used in emulation. The process information includes the process id, the parent id, process network connections, the process owner and the process name. The user information includes usernames, UIDs, groups, GIDs, passwords, and password hashes. The system information includes the operating system type, distribution, version, and patches, and the hardware architecture.

\begin{figure*}[htbp]
\centerline{\includegraphics[width=\linewidth]{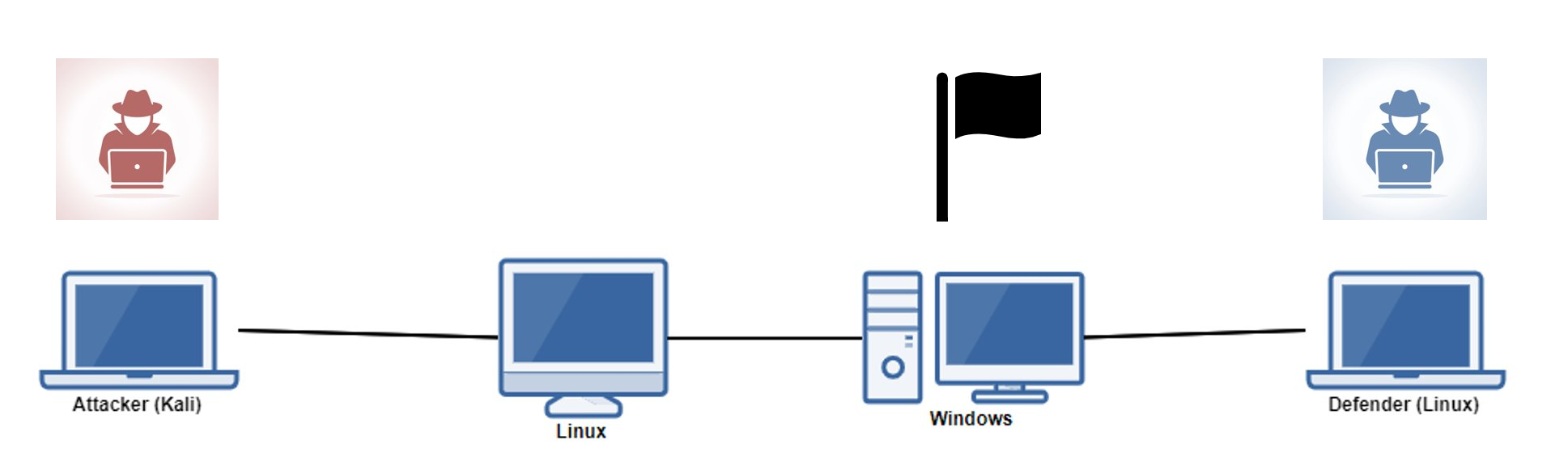}}
\caption{Diagram of CybORG Scenario.}
\label{figScenario}
\end{figure*}

\subsection{Actions}

CybORG interacts with agents via the OpenAI interface \cite{brockman2016openai}. This interface presents an observation and action space to the agent, and the agent responds by selecting an action (with appropriate parameters) to perform. These actions are defined as part of the scenario, and are based on those that would be available to a cybersecurity professional in similar circumstances. In the simulator, the effect of these actions is modelled. In the emulator, these actions are fully implemented where possible. 

\subsection{Observations}

Once an agent selects an action it is performed and the agent is returned an observation object describing the new state of the system as observed by that agent. Each agent is only provided data for those parts of the scenario it could conceivably observe, depending on the reconnaissance tools available to that agent. 

The observation from CybORG is a dictionary with key-value pairs. One key in each observation, `success', indicates if the previous action performed by that agent was successful, unsuccessful or `unknown' (i.e. no information was received regarding the success of the action). The other parts of the observation are split by host id, with each value corresponding with a host id containing a dictionary that contains further observations from that host. These dictionaries divide the state further into the categories Interface, Session, User, System and Process. 

In addition to the observation, the agent is presented with a reward value and a flag indicating if the current run is complete. Once the run is complete, the reset() function returns the scenario's state to its initial state with re-randomised values where appropriate.

\subsection{Simulator}

The simulator represents the scenario as a finite state machine, where the current state represents the state of all systems and networks in the scenario. Actions use the values in the current state to determine the next state, update the values inside of the current state such that they match the next state, and then return the observable subset of this updated state to the agent. 

Simulated actions are defined by their preconditions and effects. The preconditions of an action are the state conditions that must be satisfied for the action to be successful. The effects define how the action will change the state of the environment if the action is successful.

To reduce the chance of the simulator model diverging from the behaviour of the emulator -- in particular, allowing actions to succeed where they would not in real systems -- the state includes details such as the creation or deletion of individual files or the making and breaking of network connections.

\subsection{Emulator}

The emulator currently uses Amazon Web Services (AWS) with virtual machines to create a high fidelity cyber security environment with which an agent can interact.

The emulator uses the description of the scenario from the YAML scenario files to deploy and configure a virtual network in Amazon Web Services (AWS) (as well as for tearing them down). It does this by using SSH to access a virtual gateway server in a private AWS cloud and then deploying and configuring environments using AWS's Command Line Interface (CLI) on that virtual (master) host using the following functions:

\begin{itemize}
\item Automatically creating (and deleting) IPv4 subnets based on the number of IP addresses required in each subnet according to the scenario.
\item Automatically configuring routing between subnets according to the scenario.
\item Automatically creating (and deleting) instances of static host images and assigning them to subnets according to the scenario.
\end{itemize}

Using these high-level functions, CybORG is capable of rapidly and concurrently deploying independent clusters of hosts and subnets, allowing multiple instantiations of a scenario to be run in parallel. By using IaaS it also has the advantages those offerings provide in that it is scalable, efficient and low cost. 

The CybORG Emulator can run in two different modes: pre-deployed or deployed. In the first the VMs are already running and can be interacted with as is and in the second saved VM snapshots are started before CybORG can interact with the environment.

To implement actions, the CybORG emulator uses a series of actuator objects. These connect to VMs using SSH or specialised session handlers for third party tools. Third-party tools currently used in CybORG include the Metasploit Framework \cite{maynor2011metasploit} and Velociraptor \cite{lohanathan2020live}. Other session handlers can be added as required. These actuator objects interact with security tools and systems either through APIs or terminal commands. The results of these actions are then filtered and merged to present a single observation back to an agent. This control method allows for multiple adversarial agents to act simultaneously in the environment, and will potentially allow human operators to interact within the emulator in parallel with agents.
\subsection{Summary}

Through the approach described above, CybORG is able to implement a novel capability for cyber autonomy -- the ability to train and test the same agent, using the same body of code, at differing levels of fidelity. Agents can be trained, using standard learning approaches, in the simulator. The effectiveness of these agents can then be validated on virtual infrastructure. In the next section, we describe a scenario implemented within CybORG for the development of an autonomous penetration testing agent using RL.

\section{Experimentation}
% !TeX spellcheck = en_GB
\subsection{Scenario}

Our initial test scenario is shown in Figure \ref{figScenario}. While only having three hosts for an attacker to access, it is of sufficient complexity and detail to capture examples of most attacker behaviours within the cyber killchain, as first defined in \cite{hutchins2011intelligence}.%many aspects of a cyber security environment. 

The scenario consists of 3 hosts split into 2 subnets. The attacker host runs a Metasploit server that allows the agent to perform all parts of the killchain. The attacker is in its own subnet, which does not block any traffic and allows packets to be sent to hosts in the other subnet. 

The other two hosts are the Internal and Gateway hosts that sit in the Internal subnet. The Gateway host is an Ubuntu 18 host that has an open SSH port. The Internal host is a Windows 2008 server with open SSH and SMB ports. The SSH services have sufficiently simple sets of credentials for the server to be vulnerable to a brute force attack -- the Gateway host has the username “pi” and password “raspberry”, and the Internal host has the username “vagrant” and password “vagrant”. The SMB service is vulnerable to the Eternal Blue (MS17-010) exploit. 

There is a single agent in the scenario that is a red agent. The goal of the red agent is to get a session on the Internal host as the System user, and thereby have full access to the internal host's file system. In order to achieve this goal, the agent must be able to perform reconnaissance on hosts, exploit these hosts, establish meterpreter sessions and use a meterpreter session to pivot between machines. An effective red agent will be able to select appropriate actions, with correct parameters, to complete each of these sub-goals in the proper order.

To model these actions, and then implement these actions in the Emulator, we use the Metasploit Framework and Meterpreter as the actuators and sensors for the agent. The actions available to the red agent are: SSH Bruteforce, Portscan, Pingsweep, Upgrade to Meterpreter, IPConfig, MS17-010-PSExec, Autoroute, and Sleep. Each of these actions has associated parameters, for which the agent must learn the correct values.

The red agent receives a large reward for starting a session on the Internal host as the System user. The agent also receives moderate rewards for gaining user sessions on the Internal and Gateway hosts, and minor rewards for discovering new information about the network. For this simple scenario the reward did not take into account any action costs.

\subsubsection{Agent Training Methodology}

In order to demonstrate the transferability of training from the simulator to the emulator, an RL agent is initially trained on the simulator then run on the emulator. 

This RL agent uses a Deep Q-Network \cite{Mnih2015} to learn a policy which maps the current state of the environment to the discounted rewards for each action. Learning this policy, over a large series of training runs, allows the agent to select an action at each step that will produce the highest reward. 

For this scenario, the agent will be unable to learn an effective policy without remembering key features of the scenario. This memory is implemented through a Long Short Term Memory, as described in \cite{DBLP:journals/corr/HausknechtS15}. 

For approaches such as Deep RL the observation and action space provided by CybORG require modification. We construct a suitable observation space for the agent using a wrapper around CybORG. This wrapper turns the elements of the observation into a single vector of floating point numbers with a fixed size. This is the input required for Deep Q-learning as used here. We also use a wrapper that takes a vector of integers from the RL agent and converts them into an action (and parameters, if required) that can be performed in CybORG. 

To select the action and parameters, our agent uses a Neural Network that has a stem that takes in an observation, constructs a feature vector from that observation, interprets it through an LSTM module, then splits into branches which each output a Q-value for either an action or a parameter. We provide the already determined parameters which the agent uses to mask its own parameter selection. This masking prevents the agent from randomly or intentionally selecting a parameter that it has not yet encountered. This improves the realism of the scenario and the convergence rate of the agent.

\subsection{Results}

We trained each RL agent for up to 2500 iterations in the simulator. This was selected as it was well above the average number of iterations for a successful run in initial testing, but short enough for multiple training runs to be conducted on our hardware in a reasonable amount of time. Each iteration took a maximum of 20 steps. An iteration was stopped if an agent could get access to the Windows host using the System account. If an agent was able to achieve this within 10 steps, then that agent was deemed to be successful and the training run was finished; otherwise the agent was deemed not successful. A minimum of 7 steps was required for an agent to get System access on the Windows host. With these parameters, we were able to generate an effective red agent on every training run.

To test the ability of these agents to transfer to the emulator, we took a selection of these trained agents and reran them on CybORG in emulation mode. In this case, they ran exploits using the Metasploit framework on virtual hosts within our testing network.

The initial training produced 21 independent RL agents. Each RL agent was evaluated in the emulator 10 times. The total number of successful tests was 139, giving a 66\% rate of success.

\begin{figure}[htbp]
\centerline{\includegraphics[width=\linewidth]{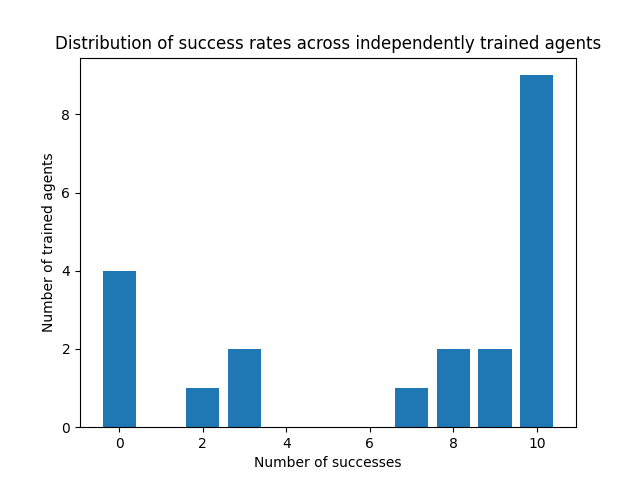}}
\caption{Distribution of Successful Runs for Trained Agents.}
\label{figGraph}
\end{figure}

Figure \ref{figGraph} shows the distribution of success rates across independently trained RL agents. Almost half of the trained agents were successful on every emulator run. Four trained agents were never successful on emulator runs, with the rest having a varying numbers of successes. 
The unsuccessful emulator runs on one hand might be explained by occasional failures to be expected in pentesting activities. The agents that consistently failed suggested that the simulation was flawed and needed refinement.
As one example, one agent was unsuccessful in an emulator run because it did not use the autoroute action to route an attack around the external firewall. The likely reason for this is that the RL agent learned a policy function that overfit to an artefact in the observation received from the simulation that was not in the observation received from the emulation. 

These results demonstrate that the underlying concept of CybORG is feasible; that a simulation can be used to train an agent which can then run effectively on virtualised infrastructure using professional security tools. While there were a significant number of failures to transfer trained agents from simulator to emulator for this scenario, these failures were sometimes associated with deficiencies in the simulator model or in the emulator's interface between the agent and the emulation environment. These deficiencies could be detected and used to refine CybORG for future training, thus demonstrating the feasibility of the approach of combining simlator and emulator in order to refine the simulator model or emulator interface. 

\section{Conclusion}
This paper introduced CybORG, a work-in-progress gym for Autonomous Cyber Operations research.

Whilst there have been ongoing developments towards simulation and emulation environments for cyber training and/or experiments, the requirements of ACO motivate an integrated design comprising emulation and simulation modes to support large scale RL across diverse scenarios.

We have made progress towards implementing this design, with the ability to spawn and play games either in simulation mode or emulation mode with cloud infrastructure. In CybORG, we can now train an RL agent in simulation then test its effectiveness in emulation. The results of this testing can alternately validate the effectiveness of the agent and act as a guide for further refinement of the simulator model or emulator interface.

CybORG or gyms like it could assist the RL research community to apply their techniques to solve cyber operations problems by providing an easy to use toolkit, a set of benchmark problems for guiding and measuring progress, and a pathway toward validation on virtual infrastructure.

\section{Future Work}

This paper shows results of red agent training. The first focus of future work is to implement the required actions, sensors and actuators to conduct the same training for a blue agent. The second focus of future work is to develop a more complex scenario with the potential for blue to employ deception.

\section*{Acknowledgments}

We would like to acknowledge Callum Baillie, Jonathon Schwartz and Michael Docking for their previous contributions to CybORG.

\section*{Availability}

Limited access to CybORG will be available for reviewers, with a public release planned once the scenario library is expanded.

\bibliographystyle{named}
\bibliography{references}
\quad
\end{document}